\RequirePackage{fix-cm}
\documentclass[smallextended]{svjour3}       
\smartqed  
\usepackage{graphicx}
\usepackage{multirow}
\usepackage{threeparttable}
\usepackage{amsmath}
\usepackage{color}
\usepackage{natbib}

\usepackage{geometry}
\usepackage{lineno,hyperref}
\begin{document}
\title{Pattern recognition in micro-trading behaviors before stock price jumps: A framework based on multivariate time series analysis
}
\subtitle{}

\titlerunning{Pattern recognition in micro-trading behaviors}        

\author{Ao Kong         \and
        Robert Azencott \and
        Hongliang Zhu \and
        Xindan Li 
}


\institute{A. Kong \at
              School of Management and Engineering, Nanjing University, Nanjing 210093, China \\
              School of Finance, Nanjing University of Finance and Economics, Nanjing 210023, China\\         
           \and
            R. Azencott \at
              Department of Mathematics, University of Houston, TX 77204, USA\\
              \and
            H. Zhu \at
            School of Management and Engineering, Nanjing University, Nanjing 210093, China \\
            \and
            X. Li* \at
              School of Management and Engineering, Nanjing University, Nanjing 210093, China \\
              \email{xdli@nju.edu.cn}
}

\date{Received: date / Accepted: date}

\maketitle

\begin{abstract}
Studying the micro-trading behaviors before stock price jumps is an important problem for financial regulations and investment decisions. In this study, we provide a new framework to study pre-jump trading behaviors based on multivariate time series analysis. Different from the existing literature, our methodology takes into account the temporal information embedded in the trading-related attributes and can better evaluate and compare the abnormality levels of different attributes. Moreover, it can explore the joint informativeness of the attributes as well as select a subset of highly informative but minimally redundant attributes to analyze the homogeneous and idiosyncratic patterns in the pre-jump trades of individual stocks. In addition, our analysis involves a set of technical indicators to describe micro-trading behaviors. To illustrate the viability of the proposed methodology, an application case is conducted based on the level-2 data of 189 constituent stocks of the China Security Index 300. The individual and joint informativeness levels of the attributes in predicting price jumps are evaluated and compared. To this end, our experiment provides a set of jump indicators that can represent the pre-jump trading behaviors in the Chinese stock market and have detected some stocks with extremely abnormal pre-jump trades.
\end{abstract}

\section{Introduction}
Large and discontinuous changes, known as jumps, are essential components of stock price dynamics \citep{Merton1976}. As perceived risk differs in small and regular price movements, they play an important role in risk measurement, option pricing, and portfolio allocation \citep{Duffie2001,Jarrow1984,Kapadia2019,Zhou2019,Chen2019}.

Among the studies related to stock price jumps, some recognize that trading behavior patterns before price jumps help analyze market stress as well as evaluate the predictability of information release, making them important for controlling market risks\citep{Bates2000,Makinen2019}. It is also point out that exploring the predictability of micro-trading indicators for price jumps is useful for evaluating the future returns and risk levels of individual stocks\citep{Kapadia2019}. However, apart from these substantial insights, empirical studies of trading behavior patterns before price jumps are scarce, perhaps because jumps, which occur instantly, often follow the micro precursors hiding in the complex noise of price dynamics.

Recently, several empirical studies, including those by \cite{Boudt2014}, \cite{Wan2017}, \cite{Jian2020} and \cite{Sun2020}, were performed to investigate liquidity dynamics before price jumps based on the high-frequency data, but few have considered the temporal information embedded in the liquidity measures. Moreover, their approaches cannot effectively compare the informativeness of different measures.

The recent development of the multivariate time series classification technique promotes innovative approaches to this problem. The feature subset selection (FSS) technique proposed by \cite{Ircio2020} is capable of exploring the temporal information on time series variables and selecting a subset of highly informative but lowly redundant attributes for multivariate time series classification.

Based on the FSS technique of \cite{Ircio2020}, we propose a novel research methodology within which to investigate the micro-trading dynamics before price jumps. Apart from the liquidity measures in the related literature, we also involve a number of technical indicators to describe intraday trading behaviors, the idea of which is inspired by previous work of \cite{Kong2020}. Time series-based mutual information, as a key point of the FSS technique, is used to evaluate the abnormality embedded in the attributes related to the occurrence of price jumps. The FSS technique is then used to select a subset of the attributes, called jump indicators, that can efficiently represent the abnormal patterns of trading behaviors before price jumps for all stocks. To this end, our method identifies the common and idiosyncratic patterns of the stocks using these indicators and time-series based clustering techniques. As an application example, the whole methodology was applied to the high-frequency data of the constituent stocks of the China Security Index 300.

The main contributions of this study are fourfold. First, we propose a novel framework based on multivariate time series analysis to study trading behavior patterns before stock price jumps. Compared with existing methods, our approach takes into account the temporal information when evaluating the informativeness of trading-related attributes to better explore the duration and level of the abnormality and compare attributes. Moreover, our approach can evaluate the joint power of the candidate attributes to recognize abnormal patterns as well as select a set of jump indicators for analyzing the trading-related idiosyncratic patterns of individual stocks. Second, in addition to commonly used liquidity measures, our study proposes using a set of technical indicators that have low mutual dependencies with each other as well as with liquidity measures from the perspective of time series information. The whole set of candidate attributes allows us to assess trading behaviors from various perspectives. Third, based on our approach, we find that in Chinese stock market although a large number of abnormalities are present shortly before the occurrence of jumps, some start much earlier. This finding complements the literature on abnormal trading movements from the perspective of time series analysis. Fourth, our study provides a generic set of jump indicators for recognizing abnormal patterns in the Chinese stock market, and have detected some stocks with extremely abnormal trading behaviors before price jumps.

The remainder of this paper is organized as follows. Section 2 provides the preliminaries of the new multivariate time series classification technique. Section 3 details our proposed research framework. Section 4 shows the experimental results and Section 5 concludes.

\section{Literature review}
\subsection{Liquidity dynamics before price jumps}
There is limited literature on the micro-trading behavior patterns before stock price jumps. As news announcements can force price jumps, early related research concentrates more on the effect of news announcement on the market and investigates liquidity dynamics around microeconomic or firm-specific news \citep{Ranaldo2008,Lakhal2008,Jiang2011, Riordan2013,Groth2014}. While few jumps are directly associated with public news \citep{Joulin2008,Lahaye2011}, conditioning the intraday analysis on these scenarios does not exactly capture the trading behaviors of market participants before stock price jumps.

\cite{Boudt2014} recently propose a framework to investigate the liquidity dynamics before stock price jumps in the U.S. market, concentrating exactly on the ``jumping'' events. They find that liquidity measures such as trading volume, the number of trades, and the quoted and effective price spread exhibit significant abnormal movements. Following that, other researchers, such as \cite{Bedowska2016}, \cite{Wan2017}, \cite{Jian2020} and \cite{Sun2020}, conduct additional experiments on other stock markets or stock futures markets and reveal similar or different liquidity dynamics before the price jumps. Despite these different markets, they all follow similar event study approaches: after extracting the values of the liquidity measure within the window before the intraday price jumps, statistical tests are used to compare their values with those within non-jumping days averaging over all the whole time window or from minute to minute. Such approaches have two main drawbacks. On one hand, they neglect the temporal information embedded in the whole time series of liquidity measures; on the other hand, they cannot feasibly compare the informativeness of different measures.

\subsection{FSS techniques for multivariate time series classification}
FSS is a necessary procedure to reduce redundant features when dealing with high-dimensional multivariate classification. When the variables are time series, this is even more important because processing a large volume of time series is neither efficient nor effective \citep{Lee2012a}. The FSS technique can be categorized into three groups: wrapper methods, filter methods, and embedded methods \citep{Das2018}. Among these, filter methods are the most popular because they do not depend on a specific classifier, which is less prone to overfitting and cheaper to compute \citep{Vergara2014}.

The majority of filter FSS methods based on multivariate time series are dedicated to solving forecasting problems and fewer relate to classification problems \citep{Fang2015, Jovic2017, Han2015, Yoon2005, He2019}. Despite their different methodologies, most need to transform the original time series into different representations, raising the issue of losing the interpretability of or information on the original data. Before the recent study by \cite{Ircio2020}, only \cite{Saikhu2019} have provided an approach to select time series subsets without transforming the original data. However, this method is unsuitable for continuous time series variables.

\cite{Ircio2020} describes an FSS technique that can select informative subsets from continuous time series features for classification without needing to transform them into different representations. The key point of this technique is to measure the informativeness and mutual dependency of the time series by mutual information. It can explore both the linear and non-linear patterns of the data, which is more suitable for complex financial systems\citep{Bedowska2021,Leong2016}. Thus, owing to the above advances, the FSS method by \cite{Ircio2020} is adopted in our study to evaluate the informativeness levels of the micro-trading-related time series and explore their joint power to explain the trading patterns of individual stocks before price jumps.

 \section{Techniques for multivariate time series classification}
Consider $P$ random variables $TS^{(1)}, TS^{(2)}, \cdots, TS^{(P)}$, whose realizations are time series. Assume we have a set of samples $\{X_1, X_2, \cdots, X_N\}$, where $X_n = (ts_n^{(1)}, ts_n^{(2)}, \cdots, ts_n^{(P)})$, $n \in \{1, 2, \cdots, N\}$, and $ts_n^{(p)}$ is a realization of $TS^{(p)}$, $p \in \{1,2,\cdots,P\}$. Let $C$ be a discrete classification variable, with associate labels for each of the samples $\{c_1,c_2, \cdots, c_n\}$. Although $C$ can take values in a finite set, only two class labels (1/0) are used in our study.

\subsection{Mutual information between time series}\label{MI_TSTS}
To estimate the mutual information between two time series, $TS^{(p)}$ and $TS^{(q)}$, consider a set of ``sliced'' samples $\{\tilde{X}_1, \tilde{X}_2, \cdots, \tilde{X}_N\}$, where $\tilde{X}_n=(ts_n^{(p)}, ts_n^{(q)}), n \in \{1,2,\cdots,N\}$ is a reduced sample of $X_n$ with $ts_n^{(p)}$ and $ts_n^{(q)}$ represents the realizations of $TS^{(p)}$ and $TS^{(q)}$.

Let $\xi(n)$ be the distance from a sample $\tilde{X}_n$ to its $k$th nearest neighbor, where the distance $Dist(\tilde{X}_m,\tilde{X}_n)$ between two samples $\tilde{X}_m = (ts_m^{(p)}, ts_m^{(q)})$ and $\tilde{X}_n=(ts_n^{(p)}, ts_n^{(q)})$ is defined as the maximum value along those two variables
\begin{equation}
Dist(\tilde{X}_m, \tilde{X}_n) = max\{dist(ts_m^{(p)}, ts_n^{(p)}), dist(ts_m^{(q)}, ts_n^{(q)})\},
\end{equation}
where $dist(x,y)$ is the distance between two time series, $x$ and $y$.
Then, we can count the number $\nu_{ts_n^{(p)}}$ of time series $ts_m^{(p)}(m\in\{1,2,\cdots,N\}-\{n\})$ whose distance $dist$ from $ts_n^{(p)}$ is equal to or less than $\xi(n)$, i.e.,
\begin{equation}
\nu_{ts_n^{(p)}} = |\{m|m\in\{1,2,\cdots,N\}-\{n\}, dist(ts_m^{(p)},ts_n^{(p)})<\xi(n)\}|.
\end{equation}
Similarly, replacing $p$ by $q$, we can compute $\nu_{ts_n^{(q)}}$.

Following the method by \cite{Kraskov2004} and \cite{Ircio2020}, the mutual information $I(TS^{(p)};TS^{(q)})$ between the time series $TS^{(p)}$ and $TS^{(q)}$ is computed by
\begin{equation}
I(TS^{(p)};TS^{(q)}) = \Psi(k)+\Psi(N)-\frac{1}{k}-\Big(\frac{1}{N}\sum_{n=1}^N\big(\Psi(\nu_{ts_n^{(p)}})+\Psi(\nu_{ts_n^{(q)}})\big)\Big),
\end{equation}
where $\Psi(x)$ is the digamma function, with $\Psi(x)=\Psi(x)+\frac{1}{x}$ and $\Psi(1) = -0.5772156$ (Euler--Mascheroni constant).

\subsection{Mutual information between a time series and the classification variable}\label{MI_TSC}
To estimate the mutual information between a time series $TS$ and the classification variable $C$, consider the realizations $\{ts_1, ts_2, \cdots, ts_N\}$ of $TS$ and their associated class labels $\{c_1,c_2, \cdots, c_N\}$.

Let $d(n)$ be the distance from a time series sample $ts_n$ to its $k$th nearest neighbors within the subset belonging to the $c_n$ class. Then, we can count the number $\nu_{ts_n}$ of time series $ts_m, m\in\{1,2,\cdots,N\}-\{n\}$ whose distance from $ts_n$ is equal to or less than $d(n)$, i.e.,
\begin{equation}
\nu_{ts_n} = |\{m|m\in\{1,2,\cdots,N\}-\{n\}, dist(ts_m,ts_n)<d(n)\}|.
\end{equation}

Following the method by \cite{Ross2014} and \cite{Ircio2020}, the mutual information $I(TS; C)$ between the time series $TS$ and classification variable $C$ is computed by
\begin{equation}
I(TS;C) = \Psi(k)+\Psi(N)-\Big(\frac{1}{N}\sum_{n=1}^N\big(\Psi(\nu_{c_n})+\Psi(\nu_{ts_n})\big)\Big),
\end{equation}
where $\Psi(x)$ is the digamma function as above and $\nu_{c_n}$ is the number of time series samples whose class labels are $c_n$.

\subsection{Mutual information-based FSS for multivariate time series classification}\label{mRMRFS}
An optimal set of features should contain a list of highly informative and minimally redundant features. In our study, the minimum-Redundancy Maximum-Relevancy (mRMR) FSS algorithm combined with mutual information, as used by \cite{Ircio2020}, is used to select the optimal set of time series features . The goal of mRMR is to select a subset of features that maximize the following score function:
\begin{equation}
J(s) = \sum_{s\in S}I(TS^{(s)};C)-\frac{1}{2|S|}\sum_{s\in S}\sum_{\substack{q\in S \\ q \neq s}} I(TS^{(s)};TS^{(q)}),
\end{equation}
where $S$ is the index of the selected features.
To achieve this goal, we apply a forward selection search strategy as by \cite{Meyer2008}: in the first step, the feature that has the largest mutual information with the class label (i.e., that with the highest discriminating power) is selected; then, in each of the following steps, when the list of $S'$ features are selected, mRMR ranks all the features $TS^{(q)}$ that are not selected according to $I(TS^{(s)};TS^{(q)})-\sum_{q\in S'}I(TS^{(s)};TS^{(q)})$ and selects the top ranked one. Meanwhile, in each step, the value of the score function is computed and the set of features is finally determined when the score function is maximized.

\section{Proposed framework}
\begin{figure}[htbp]
\begin{center}
\includegraphics[scale=0.2]{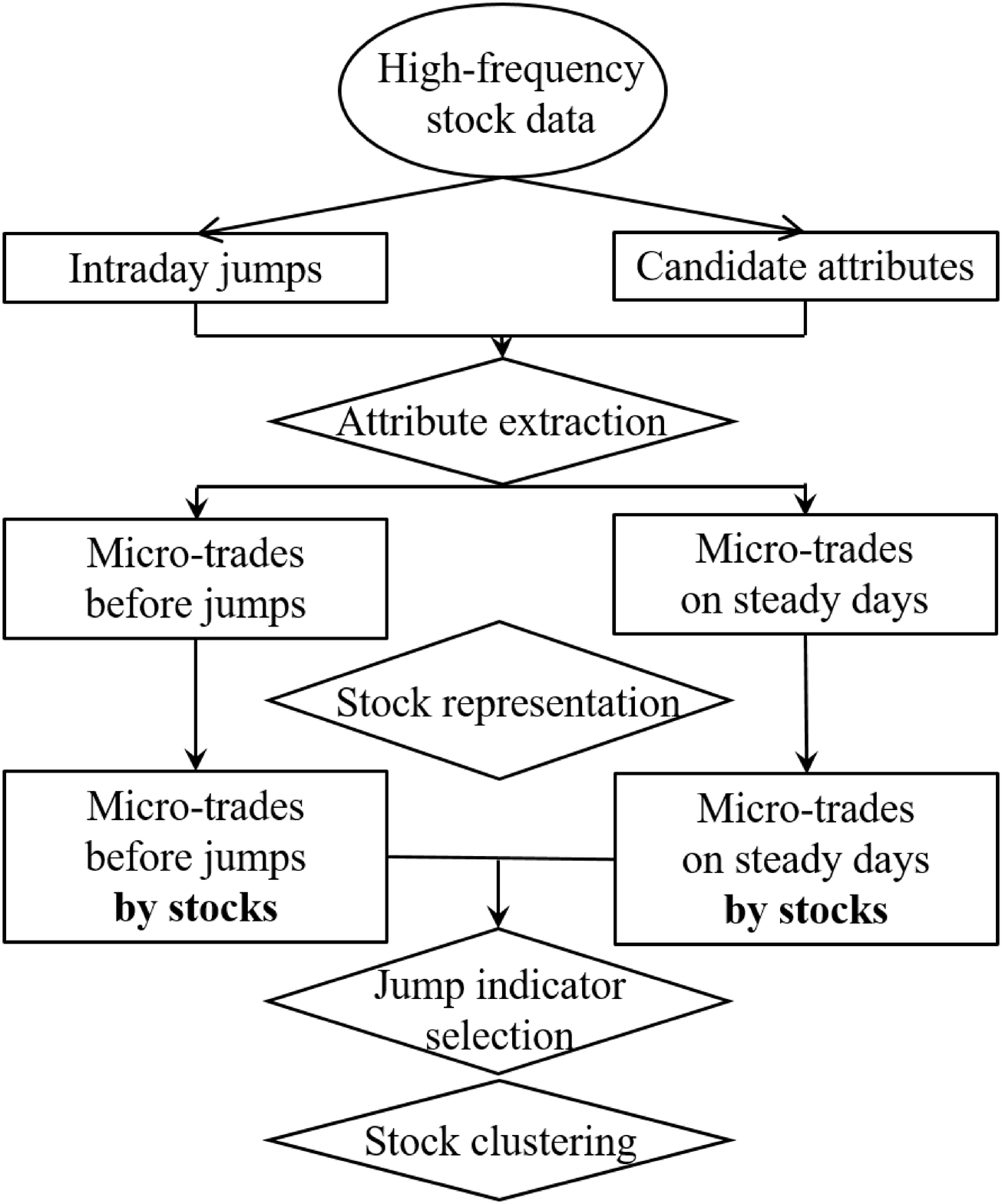}
\caption{Flowchart of our proposed framework}
\end{center}\label{fig:framework}
\end{figure}

Our research methodology uses the high-frequency data in the stock market, which can be the tick-by-tick data as used by \cite{Boudt2014} or the level-2 data as used by \cite{Kong2020}. Figure \ref{fig:framework} shows the flowchart of our proposed methodology. The intraday jumps and candidate attributes are first computed based on the high-frequency data of all the stocks, and then four main steps, including attribute extraction, stock representation, jump indicator selection and stock clustering, are pipelined. These procedures are detailed in the following sections.
\subsection{Intraday jump detection}\label{jumpdetect}
To explore trading patterns before price jumps, the first step is to detect the jump components in stock price series accurately. Our study concentrates on intraday jumps. The LM test, proposed by \cite{Lee2008}, is the most widely used technique for detecting intraday jumps and thus used in our study. It divide a trading day equally into  $\Delta t$-minute intervals, and then determines whether there is a jump in each interval.

Suppose there are $M$ intervals in a day. The LM test uses the statistic $L_i = \frac{|r_i|}{\hat{\sigma}_i}$ to examines the presence of a jump in the $i$th interval, where $r_i$ is the log return in this interval and $\hat{\sigma}^2_i =\frac{1}{K-2}\sum_{j=i-K+2}^{i-1}|r_j-1||r_j|$ is the estimated instantaneous volatility, computed by the realized bi-power variation of the returns in the previous $K-2$ time intervals. One can reject the null hypothesis that no jump is present in this interval if
\begin{equation}
\frac{L_i-C_{MT}}{S_{MT}}>-\log(-\log(1-\alpha)),
\end{equation}
where $\alpha$ is the significance level, $T$ is the total number of days. $C_{MT}$ and $S_{MT}$ are only functions of the total number of intervals $MT$ and their computation can be found in \cite{Lee2008}.

The use of 5-minute intervals to detect intraday jumps is popular in literature \citep{Bollerslev2011, Liu2015, Wan2017}. Such intervals represent a trade-off between maximizing statistical power and minimizing the effect of microstructure noise \citep{Caporin2017}.

\subsection{Candidate attributes to describe micro-trading behaviors}\label{features}
Table \ref{tab:indicator} lists the candidate attributes we use to describe micro-trading behaviors, which can be divided into liquidity measures and technical indicators. This set of attributes was processed into tabular data to fit into traditional machine learners for jump prediction in our previous study \citep{Kong2020}, while in this study, we use the time series of these attributes to study micro-trading patterns. The attributes are computed based on high-frequency transaction data, where their formulas can be found in \cite{Kong2020}.

In our methodology, the sequence the attributes are computed in a 5-minute frequency in each trading day. We compute the liquidity measures within each 5-minute interval to evaluate market quality during that short period. Though the level-2 data, which is updated every 3 seconds, is used by \cite{Kong2020}, similar computation can be performed on tick-by-tick data as used by \cite{Boudt2014}. The technical indicators are computed at the end of each 5-minute interval to describe recent market trends, which only need the 5-minute sampled price and volume information regardless of the underlying frequency of the transaction data. Of the 18 technical indicators, 12 require the parameters of lagged periods (the number of lagged intervals), which are set to 5 and 20 to take account of the trends within shorter and longer periods, respectively, providing 30 technical indicators. To this end, the trading behaviors within a trading day are described by 40 attribute series, which are computed for each interval of a day.

\begin{table}[htbp]
\centering
\caption{Candidate attributes used to describe micro-trading behaviors}\label{tab:indicator}
\begin{tabular}{lllll}
\hline\noalign{\smallskip}
     \multicolumn{2}{c}{Liquidity measures}     &\multicolumn{2}{c}{Technical indicators}\\
\noalign{\smallskip}\hline\noalign{\smallskip}
  Return                      &$r$     &Price rate of change        &$PROC(q)$\\
  Number of trades            &$K$     &Volume rate of change          &$VROC(q)$\\
  Trading volume              &$V$     &Moving average of price       &$MA(q)$ \\
  Trading size                &$S$     &Exponential moving average of price &$EMA(q)$\\
  Trade imbalance             &$TI$     &Bias to MA           &$BIAS(q)$\\
  Depth imbalance             &$DI$    &Bias to EMA            &$EBIAS(q)$\\
  Quoted spread               &$QS$    &Price oscillator to MA &$OSCP(q)$\\
  Effective spread            &$ES$     &Price oscillator to EMA &$EOSCP(q)$\\
  Realized volatility         &$RV$     &Fast stochastic \%K     &$fK(q)$\\
  Cumulative return           &$R$    &Fast stochastic \%D     &$fD(q)$\\
                              &               &Slow stochastic \%D      &$sD(q)$\\
                             &               &Commodity channel index         &$CCI(q)$\\
                             &               &Accumulation/Distribution oscillator  &$ADO$\\
                             &               &True range            &$TR$\\
                             &               &Price and volume trend              &$PVT$\\
                             &               &On balance volume         &$OBV$\\
                             &               &Negative volume index     &$NVI$\\
                             &               &Positive volume index       &$PVI$\\
\noalign{\smallskip}\hline
\end{tabular}
\begin{tablenotes}
\item In the technical indicators, $q$ represents the parameters of the lagged period.
\end{tablenotes}
\end{table}

\subsection{Attribute extraction}\label{attriextract}
To explore the abnormal patterns of the 40 attributes before price jumps, our study compares the dynamics of the attributes between the jumping and non-jumping samples. A jumping sample corresponds to a jumping interval, which is described by 40 attribute series extracted within a prior 4-hour window. According to existing literature, a 4-hour window is sufficient to observe abnormal patterns before price jumps \citep{Boudt2014,Wan2017}. A non-jumping sample corresponds to a steady days, which is defined as days without jumps during the 5 days before and the 5 days after. It is described by the 40 attribute series extracted within the day.

Because some of the attributes are highly idiosyncratic, the liquidity measures need to be standardized to compare across different stocks, days, and intraday times\citep{Podolskij2010, Boudt2014}; for a similar reason, we also standardize the technical indicators. According to \cite{Boudt2014} and \cite{Kong2020}, there are two standardization methods. One is performed by dividing each coordinate of the time series by the median of their data at the same time of the previous 60 days and subtracting 1; the other is performed by subtracting the median of their data at the same time of the previous 60 days from each coordinate of the time series. In our study, 15 types of attributes are highly idiosyncratic. Of these, $U$, $K$, $S$, $MA$, $EMA$, $TR$, $PVT$, and $OBV$ are standardized using the first method, while $OI$, $DI$, $QS$, $ES$, $RV$, $VROC$, $NVI$, and $PVI$, which are already ratios, are standardized using the second one.

\subsection{Stock representation}\label{stockprep}
\begin{figure}[htbp]
\begin{center}
\includegraphics[scale=0.23]{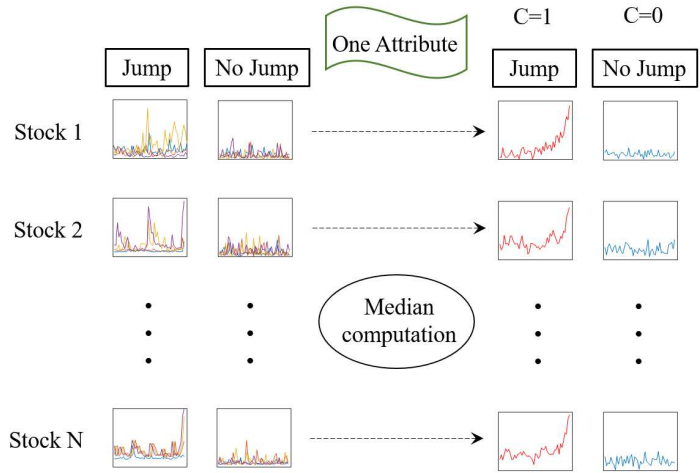}
\caption{General view of the stock representation methodology}
\end{center}\label{fig:sample}
\end{figure}

After the above data processing, we obtain a group jumping samples and a group of non-jumping samples for each stock. That is, for each attribute, there are a group of time series samples before the jumping intervals and on steady days respectively. To make a general statement about the trading patterns for each stock, we take the median values of the time series within each group for each stock. The median values are computed at each time coordinate. Figure \ref{fig:sample} illustrates the stock representation methodology.

Before price jumps, we then have $N$ time series for each attribute, where $N$ is the number of stocks. If we treat an attribute as a random variable, the $N$ time series samples can be seen as its random realizations. In other words, each stock is represented by a set of 40 time series before price jumps.

Similarly, on steady days, the median value of the time series for each attribute is computed for each stock. While the 4-hour information window under the jumping scenario can start from any time of a day, we should allow any starting point for the time series on steady days. Therefore, to perform a fair comparison, we randomize the order of the time series one by one on steady days and then compute their median values in each group. Namely, we construct a ``virtual'' time series for each attribute of each stock. We use the 40 ``virtual'' time series to represent each stock on steady days, each of which can also be recognized as a realization sample of the corresponding attribute.

For the following analysis, we label the jumping and non-jumping samples as 1 and 0, respectively. To this end, we have $2N$ binary labeled samples $\{(X_1, C_1), (X_2, C_2), \cdots, (X_{2N}, C_{2N})\}$, where $C_i\in \{1,0\}$ corresponds to the label of each sample and $X_i = (ts_i^{(1)}, ts_i^{(2)}, \cdots, ts_i^{(40)})$, $i \in \{1, 2, \cdots, 2N\}$ is represented by the 40 attribute time series.

\subsection{Jump indicator selection}\label{indselect}
Using the method described in Section \ref{mRMRFS}, we can select a set of attributes, called jump indicators that are highly related to the occurrence of price jumps. The mutual information between each attribute and label variable can be computed following the details in Section \ref{MI_TSC}, while the mutual information between the two time series is evaluated as outlined in Section \ref{MI_TSTS}. In these computations, the distance $dist(ts^{(p)},ts^{(q)})$ between the two time series is needed. We use three typical types of distances in our study, namely, the Euclidean distance, Chebychev distance, and distance based on dynamic time warping (DTW).

Different from DTW, both the Euclidean distance and the Chebychev distance treat the time series as vectors. The Euclidean distance is simply the L2-norm distance between two vectors, while the Chebychev distance takes the maximum coordinate difference between the two vectors. DTW, on the contrary, can compare two time series with varying lengths and speeds \citep{Berndt1994}. In our study, although the attribute series are always extracted with the same length, their difference in the speed of change should be considered. In general, DTW searches for an optimal match between two vectors so that the Euclidean distance between their corresponding points is minimized. Further, it has to comply with the following three rules: (i) every point in one sequence should be matched with one or more points in the other sequence, (ii) the first points and last points from both sequences should be matched, and (iii) the mapping of the points from one of the sequences to the other must be monotonically increasing and vice versa. \cite{Berndt1994} provide the detailed implementation of the DTW algorithm.

The estimation of mutual information based on the real data can be blurred by systematic errors resulting from the finite size issue \citep{Steuer2002}. To minimize this error, we correct the mutual information values by subtracting a zero baseline. Random noise should ideally have zero mutual information with any random variable. Hence, according to \cite{Steuer2002}, we randomly permutate the realizations of the two variables and compute the mutual information based on the surrogate pairs. This procedure is then repeated many times (say 100) and the average of all the mutual information values are used as the zero baseline.

\subsection{Stock clustering according to the jump indicator patterns}
While the dynamics of the trading behaviors before price jumps is described by a set of jump indicators for each stock, one might wonder how the pattern differs from stock to stock and whether some of the stocks share similar patterns before price jumps. Clustering the stocks according to their jump indicator patterns can give a fast answer to this question, as well as to detect the stocks that have idiosyncratic trading patterns before price jumps.

Clustering is the task of partitioning the samples into several groups so that the samples in one group are more similar to each other than to those in other groups. It is a widely used technique for pattern recognition and data mining problems. Many types of clustering methods exist such as connectivity-based clustering, centroid-based clustering, density-based clustering, and grid-based clustering. Different from other methods, connectivity-based clustering, also known as hierarchical clustering, does not provide a unique partitioning of the samples; instead, it generates a dendrogram showing how an extensive hierarchy of clusters merges with each other. The advantage of the method is that users can choose a set of appropriate clusters by selecting the cutoff of the inconsistency coefficient of the linkages of the dendrogram. Moreover, we need to choose the linkage criterion to compute the distance between clusters. In our study, the popular ``unweighted average linkage'' method is used.

In addition, clustering analysis can be easily dominated by extremely large attributes. Hence, to minimize the scale difference among the jump indicators, each of the indicator series $(a_1, a_2, \cdots, a_T)$ are normalized before the clustering analysis using the min-max method
\[norma_t = \frac{a_t-minv}{maxv-minv},\]
where $minv$ and $maxv$ are the minimal and maximal values of $a_t$ taken over all stocks, respectively.

\section{Application to Chinese stock market}
\subsection{Data}\label{data}
Our experiment is based on level-2 transaction data on the Chinese stock market. The stocks we consider are the constituent ones of the China Security Index 300 (CSI 300) since they are the largest and most liquid stocks in the Chinese stock market and cover about 60\% of the market's value. The study period spans January 2014 to August 2017 (896 trading days). The level-2 data, which are updated every 3 seconds, consist of the current transaction price and volume, cumulative number of trades and volume from the last record to the current one, and 10 best quotes before the current transaction. The data are downloaded from the Wind database (www.wind.com.cn). To ensure these data are adequate, stocks traded on fewer than 90\% of the 896 trading days are deleted. To minimize the effect of extreme movements of stock prices in our analysis, the stocks of companies that are newly listed, delisted, suspended, or under special treatment are removed. Finally, 189 stocks are retained for our analysis.

\subsection{Detected jumps}
Following the method in Section \ref{jumpdetect}, we detect all the intraday jumps in the 5-minute sampled price series of the 189 sample stocks. The Chinese Stock Exchange opens from 9:30 am to 11:30 am and 1:00 pm to 3:00 pm. This 4-hour trading day is divided equivalently into 48 intervals of 5 minutes. We choose the parameter $K=240$ for the detection following \cite{Lee2008} and \cite{Wan2017} and set a significance level of 1\%. Table \ref{tab:numjump} presents the number of jumps counted over all the stocks as well as their average sizes. We can see that the number of positive jumps is larger than that of negative jumps, while the average size of negative jumps is larger. This can be explained by the fact that most of the players in the Chinese stock market are retail investors that tend to chase rising prices rather than falling ones, as the market does not allow short sales; meanwhile, when prices fall, retail investors panic and short the stocks, resulting in larger negative jumps.

\begin{table}[htbp]
\centering
\caption{Statistical description of the detected jumps}\label{tab:numjump}
\begin{tabular}{lcccccccccccc}
\hline\noalign{\smallskip}
                &Positive jumps     &Negative jumps     \\
Number         &20734            &11597\\
Average return  &0.0266          &-0.0363              \\
\noalign{\smallskip}\hline
\end{tabular}
\end{table}

\subsection{Extracted attribute series}
To explore the abnormal trading behaviors of individual stocks before price jumps, we extract the attribute series within a 4-hour window before each intraday jump and within the steady days of each stock. Since our attribute are computed every 5 minutes, the length of the series for each attribute is 48.

To avoid missing time series information, we delete four types of samples from the dataset. First, we do not consider the samples whose attributes are extracted from the first 60 days of each stock. Second, because of the 10\% price change limit rule in the Chinese stock market, all the sequential intervals with limit-ups or limit-downs except the first one are deleted. Third, any intervals within suspension or a day after suspension are deleted. Fourth, because of the implementation of a circuit breaker mechanism on the Chinese stock market from April 1, 2016 to July 1, 2016, during which the stock market halted several times, the sample period between January 4, 2016 and January 8, 2016 is not considered.

\begin{figure}[htbp]
\includegraphics[scale=0.3]{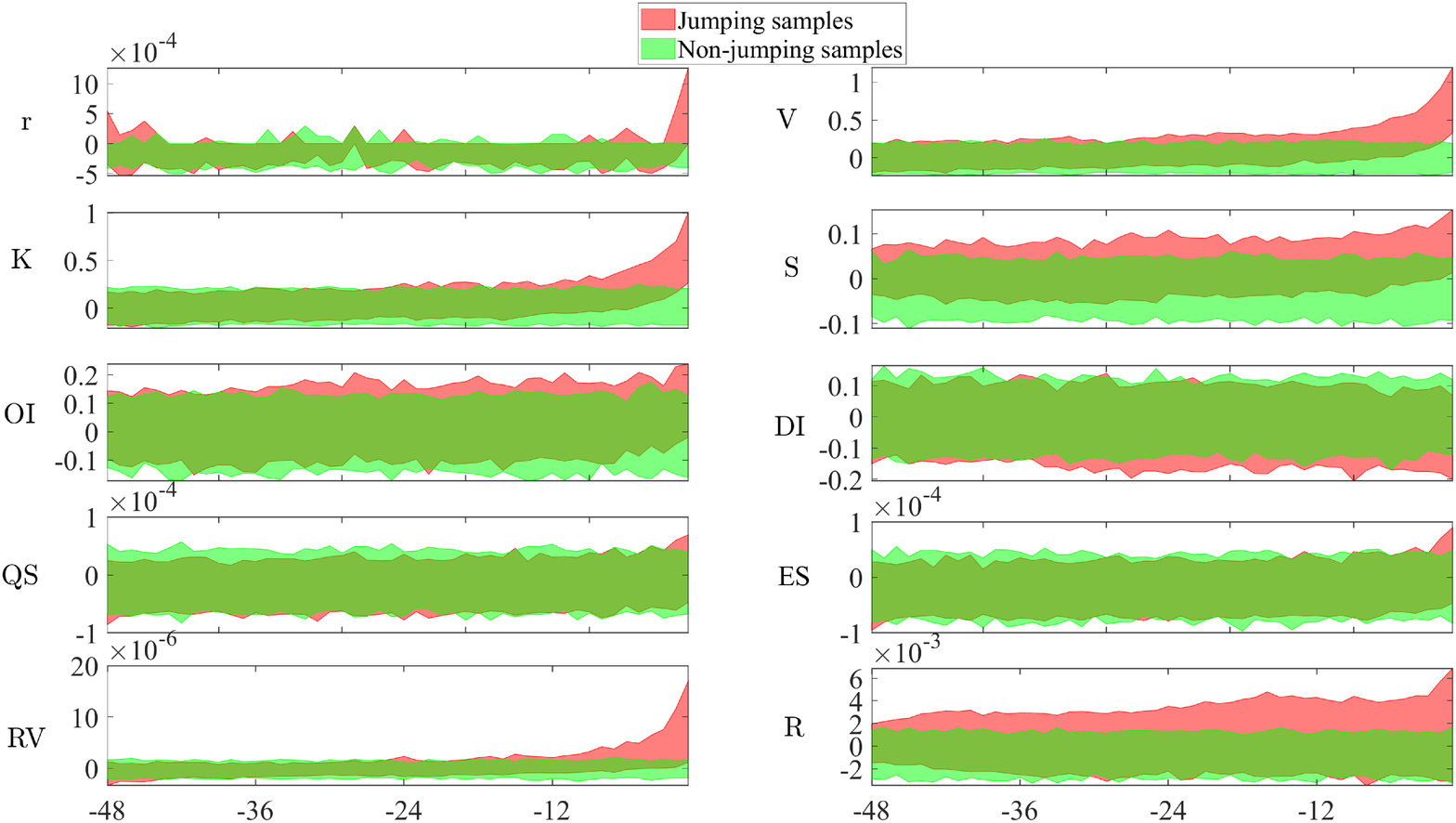}
\includegraphics[scale=0.3]{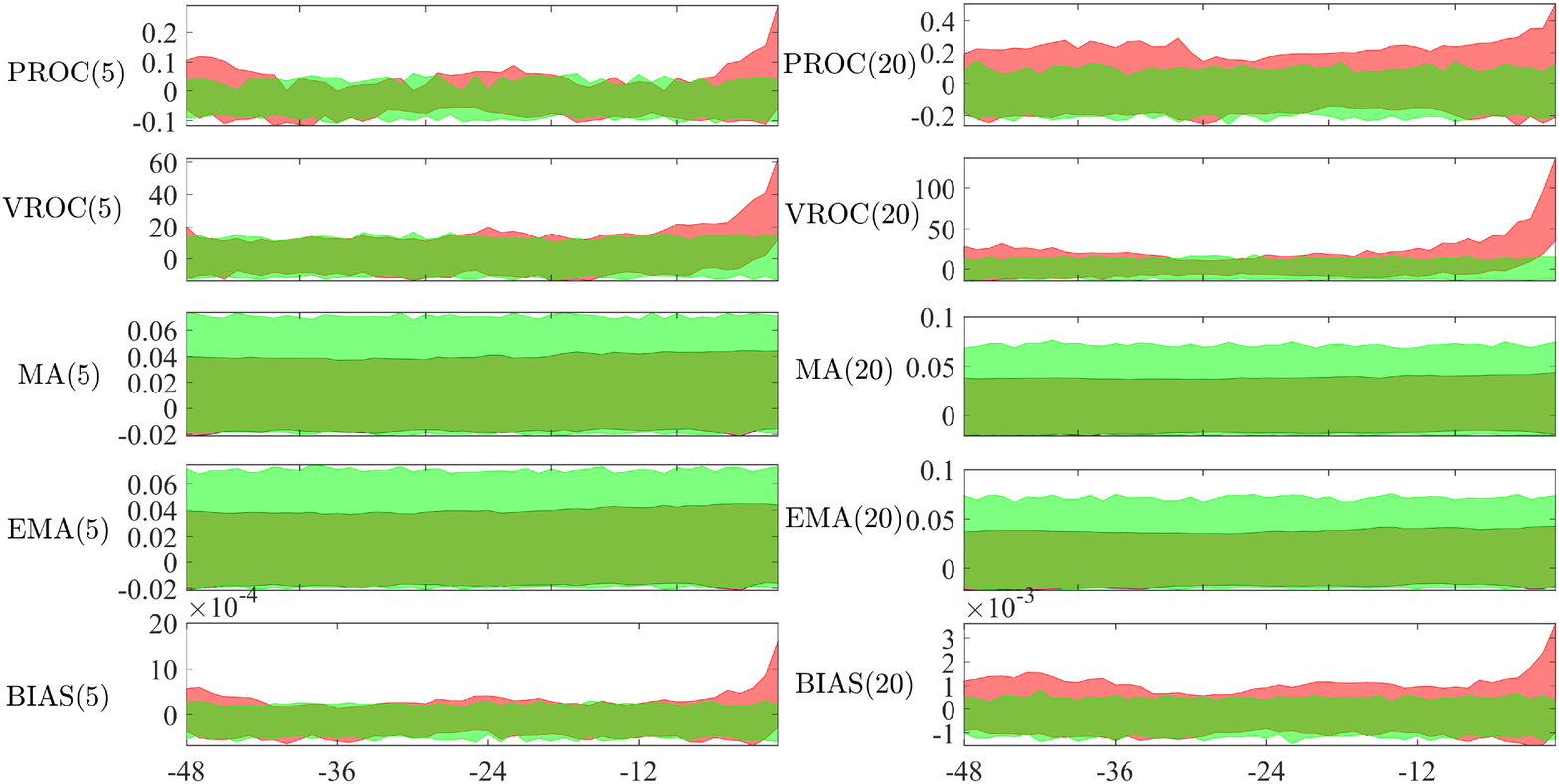}
\caption{Dynamics of the candidate attributes before price jumps and on steady days. The shaded region represents the range between the 5\% and 95\% quantiles within the 189 stock samples. The x-axis is the index of the intervals before price jumps for the jumping samples.}\label{fig:Medfignew12}
\end{figure}

\begin{figure}[htbp]
\includegraphics[scale=0.3]{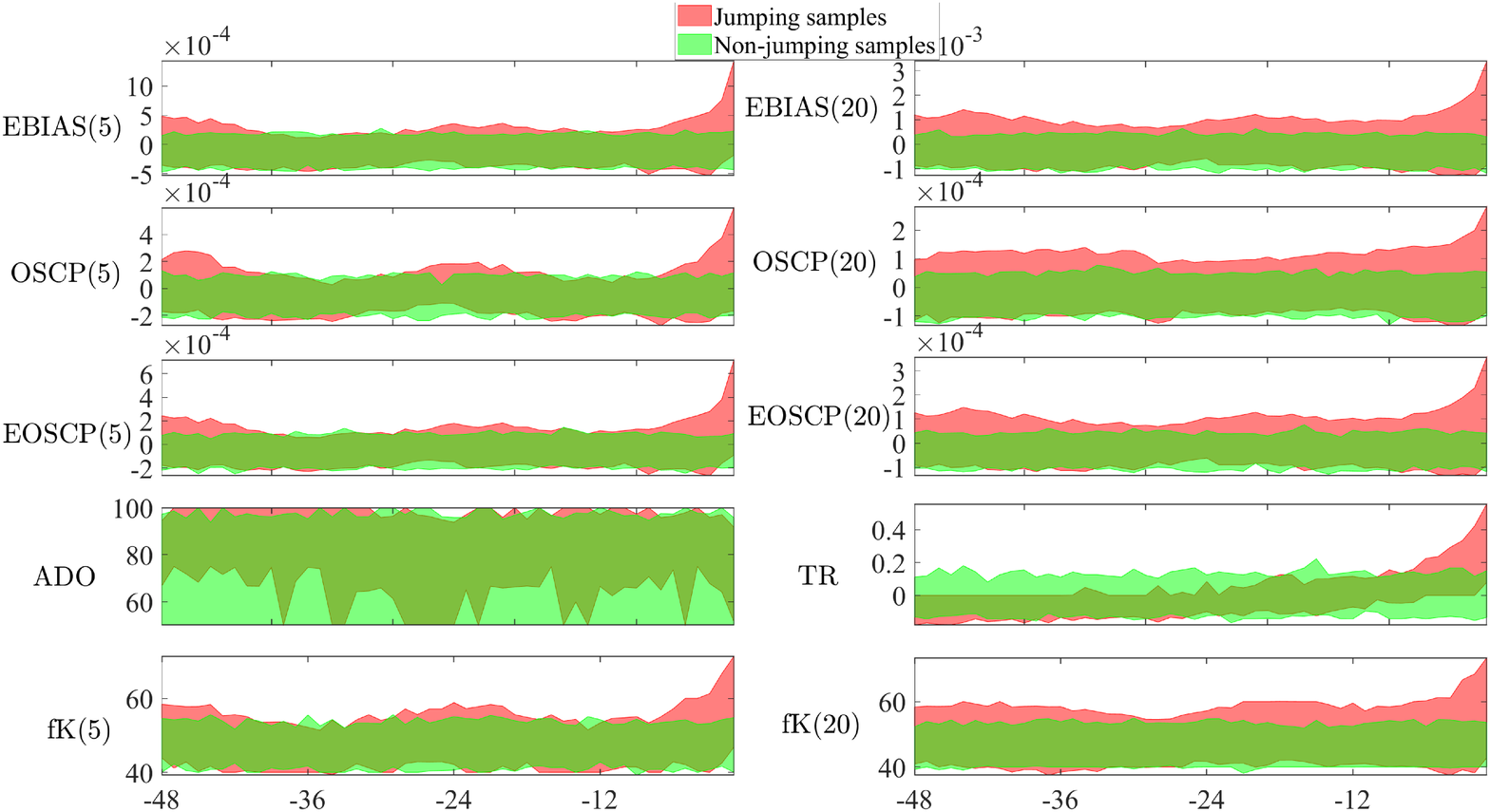}
\includegraphics[scale=0.3]{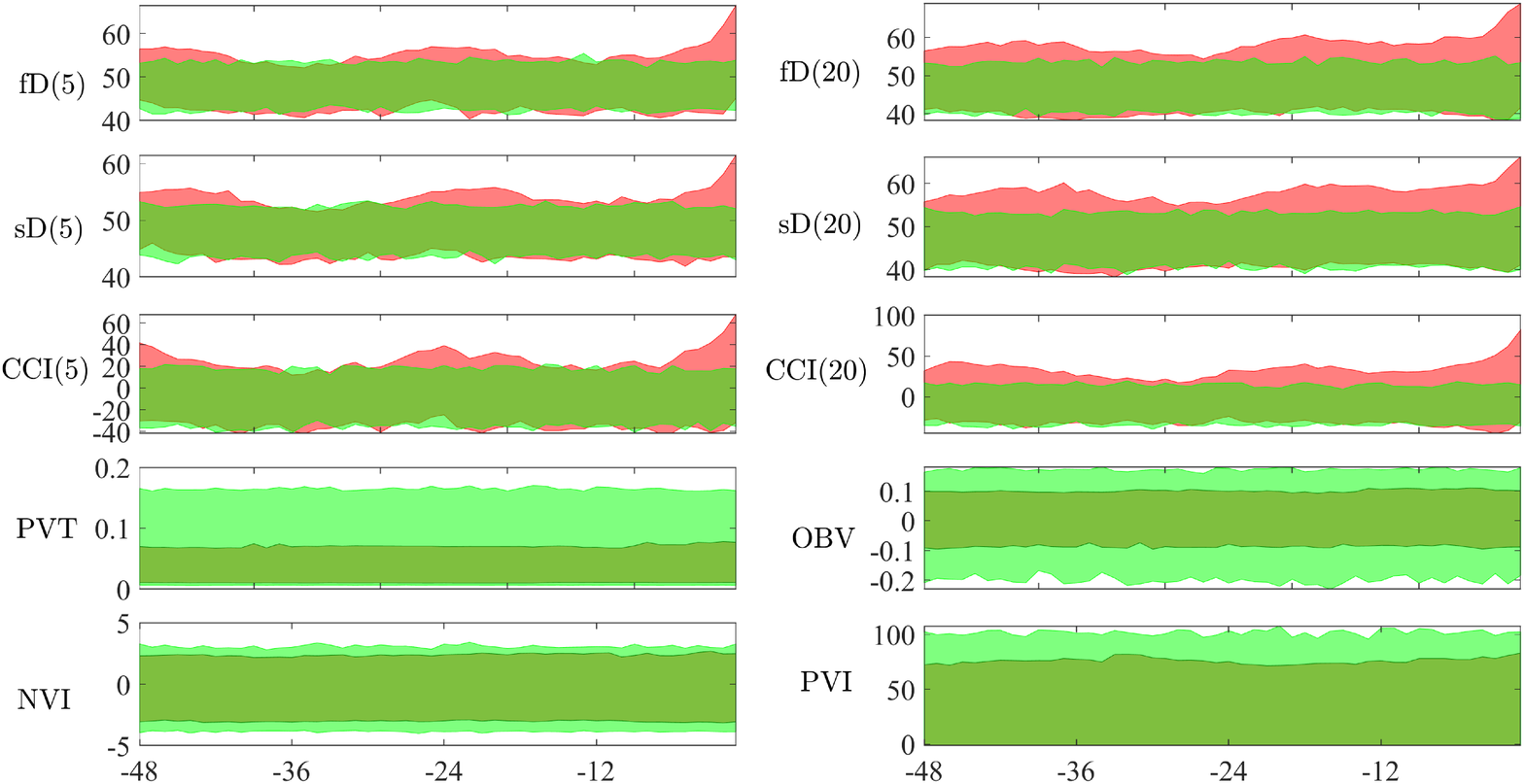}
\caption{Dynamics of the candidate attributes before price jumps and on steady days. The shaded region represents the range between the 5\% and 95\% quantiles within the 189 stock samples. The x-axis is the index of the intervals before price jumps for the jumping samples.}\label{fig:Medfignew34}
\end{figure}

After deleting the tainted samples, we still have a total of 27,816 jumps, with an average of about 150 jumps per stock. The total number steady days is 31,455, with an average of about 170 per stock. Before the jumps occur as well as on steady days, each stock is represented by the median values of the attributes, as outlined in Section \ref{stockprep}. To provide a bird's eye view of the dynamics of all the attributes in the two scenarios, Figures \ref{fig:Medfignew12}--\ref{fig:Medfignew34} plot the attribute series of the 189 samples before price jumps and on steady days respectively.

These figures show that many attributes series, including $r$, $V$, $K$, $S$, $RV$, $R$, $PROC$, $VROC$, $BIAS$,$EBIAS$ $OSCP$, $EOSCP$, $TR$, $fK$, $fD$, $sD$, and $CCI$, exhibit significant abnormality before price jumps, especially when the time window is very close the the jump occurrence. The abnormality of the liquidity measures are similar to those observed by \cite{Boudt2014} and \cite{Wan2017}, who also performed the comparison using the Mann--Whitney test. However, their conclusion is based on a point-by-point analysis, which overlooks the characteristics of the whole time series; further, it cannot compare the informativeness of the attributes. Hence, further time series-based statistical analysis needs to be conducted to explore the abnormal dynamics of the candidate attributes, for which we use the mutual information-based technique, as described next.

\subsection{Informativeness of the candidate attributes}
The mutual information between each attribute and the label variable, as defined in Section \ref{MI_TSC}, evaluates the informativeness of each attribute with respect to the occurrence of price jumps. Table \ref{tab:predpower} provides the mutual information values of the 40 attributes based on the three types of distances and 1, 3, or 5 nearest neighbors during the computation. To minimize the systematic errors resulting from the finite size issue, the zero baseline has been subtracted, as mentioned in Section \ref{indselect}.

Though there is difference between the informativeness of one attributes computed with different distances or $k$ parameter settings, it is more meaningful to compare the relative rankings of the attributes across different scenarios. Here, a two-sided Wilcoxon signed rank test is adopted to compare the ranks of the attributes between any two of the nine scenarios. At the 1\% significance level, all the tests fail to reject the null hypothesis of a zero median in the rank difference. This indicates that the choice of distance or $k$ parameters in general has a low influence when ranking the informativeness of the attributes.

\begin{table}[htbp]
\centering
\caption{Informativeness of the candidate attributes}\label{tab:predpower}
\begin{tabular}{llccccccccccc}
\hline\noalign{\smallskip}
    &&\multicolumn{3}{c}{Euclidean}  &&\multicolumn{3}{c}{Chebychev}  &&\multicolumn{3}{c}{DTW}\\
    \cline{3-5}         \cline{7-9}   \cline{11-13}
&\#Neighbors   &1 &3    &5     &&1      &3       &5         &&1      &3      &5\\
1	&	r	&	0.27 	&	0.30 	&	0.27 	&&	0.23 	&	0.21 	&	0.21 	&&	0.19 	&	0.23 	&	0.24 	\\
2	&\textbf{V}	&\textbf{0.69} 	&\textbf{0.67} 	&\textbf{0.67} 	&&\textbf{0.67} 	&\textbf{0.66} 	&\textbf{0.66} 	&&\textbf{0.67} 	&\textbf{0.68} 	&\textbf{0.68} 	\\
3	&\textbf{K}	&\textbf{0.68} 	&\textbf{0.66} 	&\textbf{0.66} 	&&\textbf{0.66} 	&\textbf{0.66} 	&\textbf{0.66} 	&&\textbf{0.68} 	&\textbf{0.67} 	&\textbf{0.67} 	\\
4	&	S	&	0.48 	&	0.49 	&	0.47 	&&	0.42 	&	0.44 	&	0.44 	&&	0.53 	&	0.50 	&	0.51 	\\
5	&	OI	&	0.33 	&	0.34 	&	0.32 	&&	0.20 	&	0.24 	&	0.25 	&&	0.40 	&	0.39 	&	0.36 	\\
6	&	DI	&	0.21 	&	0.21 	&	0.19 	&&	0.19 	&	0.15 	&	0.15 	&&	0.34 	&	0.28 	&	0.26 	\\
7	&	QS	&	0.25 	&	0.23 	&	0.22 	&&	0.24 	&	0.23 	&	0.22 	&&	0.32 	&	0.29 	&	0.26 	\\
8	&	ES	&	0.27 	&	0.29 	&	0.26 	&&	0.23 	&	0.23 	&	0.21 	&&	0.42 	&	0.37 	&	0.32 	\\
9	&\textbf{RV}	&\textbf{0.67} 	&\textbf{0.62} 	&\textbf{0.55} 	&&\textbf{0.63} 	&\textbf{0.64} 	&\textbf{0.59} 	&&\textbf{0.63} 	&\textbf{0.64} 	&\textbf{0.62} 	\\
10	&	R	&	0.50 	&	0.45 	&	0.44 	&&	0.46 	&	0.40 	&	0.41 	&&	0.58 	&	0.56 	&	0.53 	\\
11	&	PROC(5)	&	0.53 	&	0.41 	&	0.36 	&&	0.45 	&	0.39 	&	0.38 	&&	0.46 	&	0.42 	&	0.43 	\\
12	&	PROC(20)	&	0.50 	&	0.43 	&	0.37 	&&	0.46 	&	0.40 	&	0.39 	&&	0.54 	&	0.52 	&	0.51 	\\
13	&\textbf{VROC(5)}	&\textbf{0.63} 	&\textbf{0.62} 	&\textbf{0.63} 	&&\textbf{0.58} 	&\textbf{0.61} 	&\textbf{0.60} 	&&\textbf{0.66} 	& \textbf{0.67} 	&\textbf{0.66} 	\\
14	&\textbf{VROC(20)}	&\textbf{0.68} 	&\textbf{0.67} 	&\textbf{0.67} 	&&\textbf{0.65} 	&\textbf{0.66} 	&\textbf{0.66} 	&&\textbf{0.68} 	&\textbf{0.67} 	&\textbf{0.67} 	\\
15	&	MA(5)	&	0.35 	&	0.23 	&	0.17 	&&	0.35 	&	0.27 	&	0.20 	&&	0.55 	&	0.43 	&	0.35 	\\
16	&	MA(20)	&	0.36 	&	0.22 	&	0.18 	&&	0.40 	&	0.29 	&	0.22 	&&	0.58 	&	0.43 	&	0.34 	\\
17	&	EMA(5)	&	0.35 	&	0.23 	&	0.17 	&&	0.33 	&	0.25 	&	0.21 	&&	0.57 	&	0.43 	&	0.35 	\\
18	&	EMA(20)	&	0.36 	&	0.21 	&	0.15 	&&	0.35 	&	0.25 	&	0.20 	&&	0.55 	&	0.42 	&	0.33 	\\
19	&	BIAS(5)	&	0.47 	&	0.44 	&	0.43 	&&	0.49 	&	0.46 	&	0.42 	&&	0.54 	&	0.53 	&	0.52 	\\
20	&	BIAS(20)	&	0.57 	&	0.53 	&	0.51 	&&	0.54 	&	0.50 	&	0.49 	&&	0.64 	&	0.61 	&	0.60 	\\
21	&	EBIAS(5)	&	0.54 	&	0.51 	&	0.50 	&&	0.51 	&	0.48 	&	0.48 	&&	0.56 	&	0.56 	&	0.55 	\\
22	&	EBIAS(20)	&	0.60 	&	0.51 	&	0.50 	&&	0.55 	&	0.49 	&	0.46 	&&	0.63 	&	0.60 	&	0.59 	\\
23	&	OSCP(5)	&	0.35 	&	0.43 	&	0.45 	&&	0.48 	&	0.43 	&	0.42 	&&	0.46 	&	0.46 	&	0.46 	\\
24	&	OSCP(20)	&	0.47 	&	0.46 	&	0.44 	&&	0.52 	&	0.51 	&	0.46 	&&	0.55 	&	0.54 	&	0.53 	\\
25	&	EOSCP(5)	&	0.55 	&	0.51 	&	0.50 	&&	0.48 	&	0.46 	&	0.46 	&&	0.55 	&	0.56 	&	0.55 	\\
26	&	EOSCP(20)	&	0.57 	&	0.53 	&	0.50 	&&	0.53 	&	0.50 	&	0.48 	&&	0.63 	&	0.60 	&	0.58 	\\
27	&	ADO	&	0.16 	&	0.11 	&	0.11 	&&	0.08 	&	0.07 	&	0.08 	&&	0.11 	&	0.13 	&	0.12 	\\
28	&	TR	&	0.59 	&	0.57 	&	0.55 	&&	0.57 	&	0.54 	&	0.58 	&&	0.59 	&	0.60 	&	0.60 	\\
29	&	fK(5)	&	0.41 	&	0.40 	&	0.39 	&&	0.47 	&	0.42 	&	0.41 	&&	0.50 	&	0.50 	&	0.48 	\\
30	&	fK(20)	&	0.49 	&	0.48 	&	0.44 	&&	0.53 	&	0.48 	&	0.46 	&&	0.61 	&	0.61 	&	0.60 	\\
31	&	fD(5)	&	0.57 	&	0.52 	&	0.48 	&&	0.50 	&	0.45 	&	0.42 	&&	0.61 	&	0.62 	&	0.62 	\\
32	&	fD(20)	&	0.46 	&	0.48 	&	0.47 	&&	0.46 	&	0.47 	&	0.44 	&&	0.67 	&	0.66 	&	0.63 	\\
33	&	sD(5)	&	0.56 	&	0.51 	&	0.47 	&&	0.51 	&	0.43 	&	0.41 	&&	0.62 	&	0.63 	&	0.62 	\\
34	&	sD(20)	&	0.46 	&	0.40 	&	0.40 	&&	0.43 	&	0.44 	&	0.41 	&&	0.65 	&	0.63 	&	0.62 	\\
35	&	CCI(5)	&	0.53 	&	0.48 	&	0.47 	&&	0.47 	&	0.43 	&	0.42 	&&	0.63 	&	0.61 	&	0.60 	\\
36	&\textbf{CCI(20)}	&\textbf{0.59} 	&\textbf{0.53} 	&\textbf{0.52} 	&&\textbf{0.53} 	&\textbf{0.51} 	&\textbf{0.49} 	&&\textbf{0.66} 	& \textbf{0.63} 	&\textbf{0.63} 	\\
37	&	PVT	&	0.23 	&	0.13 	&	0.09 	&&	0.24 	&	0.13 	&	0.14 	&&	0.29 	&	0.18 	&	0.13 	\\
38	&	OBV	&	0.16 	&	0.11 	&	0.05 	&&	0.18 	&	0.12 	&	0.08 	&&	0.38 	&	0.20 	&	0.13 	\\
39	&	NVI	&	0.14 	&	0.08 	&	0.06 	&&	0.13 	&	0.09 	&	0.06 	&&	0.30 	&	0.19 	&	0.15 	\\
40	&	PVI	&	0.23 	&	0.10 	&	0.04 	&&	0.27 	&	0.15 	&	0.12 	&&	0.37 	&	0.20 	&	0.12 	\\
\noalign{\smallskip}\hline
\end{tabular}
\begin{tablenotes}
\item The attributes ranked within the top 10 lists under all types of distances are in bold.
\end{tablenotes}
\end{table}

The bold values in the table show that trading volume $V$, the number of trades $K$, realized volatility $RV$, the volume rate of change $VROC$, and the commodity channel index $CCI$ have very high informativeness with respect to the occurrence of price jumps. $V$, $K$, and $VROC$ are all volume-related attributes, the significant abnormality of which was also observed by both \cite{Boudt2014} and \cite{Wan2017}, indicating a demand for immediate execution before price jumps. $RV$ and $CCI$ are both volatility-related measures, related to which \cite{Wan2017} observe that volatility is significantly high before price jumps.

On the contrary, \cite{Boudt2014} and \cite{Wan2017} have also investigated the dynamics of other liquidity measures such as $QS$, $ES$, $OI$, and $DI$ before price jumps, but the relative abnormality of these measures are still unclear. However, our results show that the informativeness of $OI$ and $ES$ is moderate, while that of $DI$ and $QS$ is comparably low; nevertheless, the abnormality of the trading behaviors embedded in these attributes cannot be ignored.

\subsection{Window length of the candidate attributes}
One may also wonder whether the 4-hour window is too long to detect the abnormal dynamics of the candidate attributes, as most of the abnormality arises shortly before the price jumps occur in Figures \ref{fig:Medfignew12} and \ref{fig:Medfignew34}. Therefore, further investigation based on time series analysis from the perspective of information theory is necessary. Similar to in Table \ref{tab:predpower}, we compute the mutual information between each attribute using the class label variable, but shrink the window to 3 hours, 2 hours, 1 hour, and or 30 minutes (i.e., 36, 24, 12, and 6 intervals). That is, we evaluate the informativeness of each attribute in shorter periods before price jumps.

There is no gold standard to choosing the type of distance or $k$ parameter in $k$-nearest neighbor method \citep{Ircio2020}; however, as verified in Table \ref{tab:predpower}, the comparison of the attributes changes little when using different parameters. To be concise, we only present the results based on the Euclidean distance and 3-nearest neighbor method, as shown in Table \ref{tab:changepower}.

\begin{table}[htbp]
\centering
\caption{Informativeness of the candidate attributes in different sized windows}\label{tab:changepower}
\begin{tabular}{llccccccccccc}
\hline\noalign{\smallskip}
&Prior intervals &-48   &-36    &-24     &-12      &-6       \\
\hline
1	&	r	&	0.30 	&	0.33 	&	0.32 	&	0.31 	&	0.27 	\\
2	&	U	&	0.67 	&	0.66 	&	0.66 	&	0.65 	&	0.65 	\\
3	&	K	&	0.66 	&	0.65 	&	0.64 	&	0.63 	&	0.60 	\\
4	&	S	&	0.49 	&	0.48 	&	0.49 	&	0.49 	&	0.50 	\\
5	&	OI	&	0.34 	&	0.34 	&	0.31 	&	0.32 	&	0.33 	\\
6	&\textbf{DI}	&\textbf{0.21} 	&\textbf{0.18} 	&\textbf{0.18} 	&\textbf{0.12} 	&\textbf{0.14} 	\\
7	&\textbf{QS}	&\textbf{0.23} 	&\textbf{0.20} 	&\textbf{0.21} 	&\textbf{0.17} 	&\textbf{0.12} 	\\
8	&\textbf{ES}	&\textbf{0.29} 	&\textbf{0.23} 	&\textbf{0.24} 	&\textbf{0.23} 	&\textbf{0.18} 	\\
9	&\textbf{RV}	&\textbf{0.62} 	&\textbf{0.62} 	&\textbf{0.60} 	&\textbf{0.54} 	&\textbf{0.52} 	\\
10	&	R	&	0.45 	&	0.43 	&	0.41 	&	0.41 	&	0.39 	\\
11	&	PROC(5)	&0.41 	&	0.42 	&	0.45 	&	0.41 	&	0.37 	\\
12	&	PROC(20)&0.43 	&0.46 	&0.43 	&	0.39 	&	0.38 	\\
13	&	VROC(5)	&0.65 	&0.64 	&0.65 	&	0.65 	&	0.65 	\\
14	&	VROC(20)&0.67 	&0.67 	&0.66 	&	0.65 	&	0.65 	\\
15	&\textbf{MA(5)}	&\textbf{0.23} 	&\textbf{0.24} 	&\textbf{0.21} 	&\textbf{0.14} 	&\textbf{0.09} 	\\
16	&\textbf{MA(20)}	&\textbf{0.22} 	&\textbf{0.12} 	&\textbf{0.09} 	&\textbf{0.08} 	&\textbf{0.03} 	\\
17	&\textbf{EMA(5)}	&\textbf{0.23} 	&\textbf{0.21} 	&\textbf{0.16} 	&\textbf{0.10} 	&\textbf{0.09} 	\\
18	&\textbf{EMA(20)}	&\textbf{0.21} 	&\textbf{0.13} 	&\textbf{0.14} 	&\textbf{0.06} 	&\textbf{0.06} 	\\
19	&	BIAS(5)	&	0.44 	&	0.48 	&	0.51 	&	0.51 	&	0.48 	\\
20	&	BIAS(20)	&	0.53 	&	0.53 	&	0.54 	&	0.52 	&	0.53 	\\
21	&	EBIAS(5)	&	0.51 	&	0.54 	&	0.54 	&	0.52 	&	0.50 	\\
22	&	EBIAS(20)	&	0.51 	&	0.53 	&	0.54 	&	0.52 	&	0.53 	\\
23	&	OSCP(5)	&	0.43 	&	0.40 	&	0.51 	&	0.44 	&	0.41 	\\
24	&	OSCP(20)	&	0.46 	&	0.46 	&	0.41 	&	0.41 	&	0.43 	\\
25	&	EOSCP(5)	&	0.51 	&	0.53 	&	0.55 	&	0.54 	&	0.50 	\\
26	&	EOSCP(20)	&	0.53 	&	0.53 	&	0.54 	&	0.51 	&	0.53 	\\
27	&	ADO	&	0.11 	&	0.09 	&	0.08 	&	0.11 	&	0.12 	\\
28	&	TR	&	0.57 	&	0.53 	&	0.55 	&	0.51 	&	0.51 	\\
29	&	fK(5)	&	0.40 	&	0.44 	&	0.47 	&	0.47 	&	0.43 	\\
30	&	fK(20)	&	0.48 	&	0.48 	&	0.51 	&	0.48 	&	0.47 	\\
31	&	fD(5)	&	0.52 	&	0.50 	&	0.51 	&	0.48 	&	0.45 	\\
32	&\textbf{fD(20)}	&\textbf{0.48} 	&\textbf{0.48} 	&\textbf{0.44} 	&\textbf{0.42} 	&\textbf{0.40} 	\\
33	&\textbf{sD(5)}	&\textbf{0.51} 	&\textbf{0.51} 	&\textbf{0.48} 	&\textbf{0.43} 	&\textbf{0.41} 	\\
34	&	sD(20)	&	0.40 	&	0.37 	&	0.36 	&	0.38 	&	0.38 	\\
35	&	CCI(5)	&	0.48 	&	0.51 	&	0.52 	&	0.52 	&	0.49 	\\
36	&	CCI(20)	&	0.53 	&	0.51 	&	0.52 	&	0.50 	&	0.50 	\\
37	&	PVT	&	0.13 	&	0.13 	&	0.11 	&	0.12 	&	0.14 	\\
38	&	OBV	&	0.11 	&	0.09 	&	0.08 	&	0.03 	&	0.05 	\\
39	&	NVI	&	0.08 	&	0.08 	&	0.07 	&	0.03 	&	0.06 	\\
40	&	PVI	&	0.10 	&	0.09 	&	0.11 	&	0.06 	&	0.05 	\\
\noalign{\smallskip}\hline
\end{tabular}
\begin{tablenotes}
\item These mutual information values are computed based on the Euclidean distance and 3-nearest neighbor method. The attributes whose informativeness decreases with comparably larger rates as the information interval shrinks are in bold.
\end{tablenotes}
\end{table}

According to the definition of mutual information in Sections \ref{MI_TSTS} and \ref{MI_TSC}, if the significant abnormality of an attribute arises shortly before the occurrence of price jumps, its informativeness within a shorter period should ideally differ little than those within the 4-hour window. Table \ref{tab:changepower} shows that the informativeness of most attributes, especially the highly informative ones, change little as the window size shrinks from 48 intervals (4 hours) to six intervals (30 minutes). This indicates that many of the significant abnormalities indeed arise only shortly before the occurrence of price jumps, which is consistent with previous findings \citep{Boudt2014,Wan2017, Bedowska2016}. However, some attributes still achieve much lower informativeness within shorter windows, such as $DI$, $ES$, $QS$, $RV$, $MA$, $EMA$, $fD$, and $sD$. This implies that the abnormal movements of these attributes start much earlier and last for longer until the occurrence of price jumps.

\subsection{Jump indicators for pattern recognition}
\begin{figure}
\includegraphics[scale=0.48]{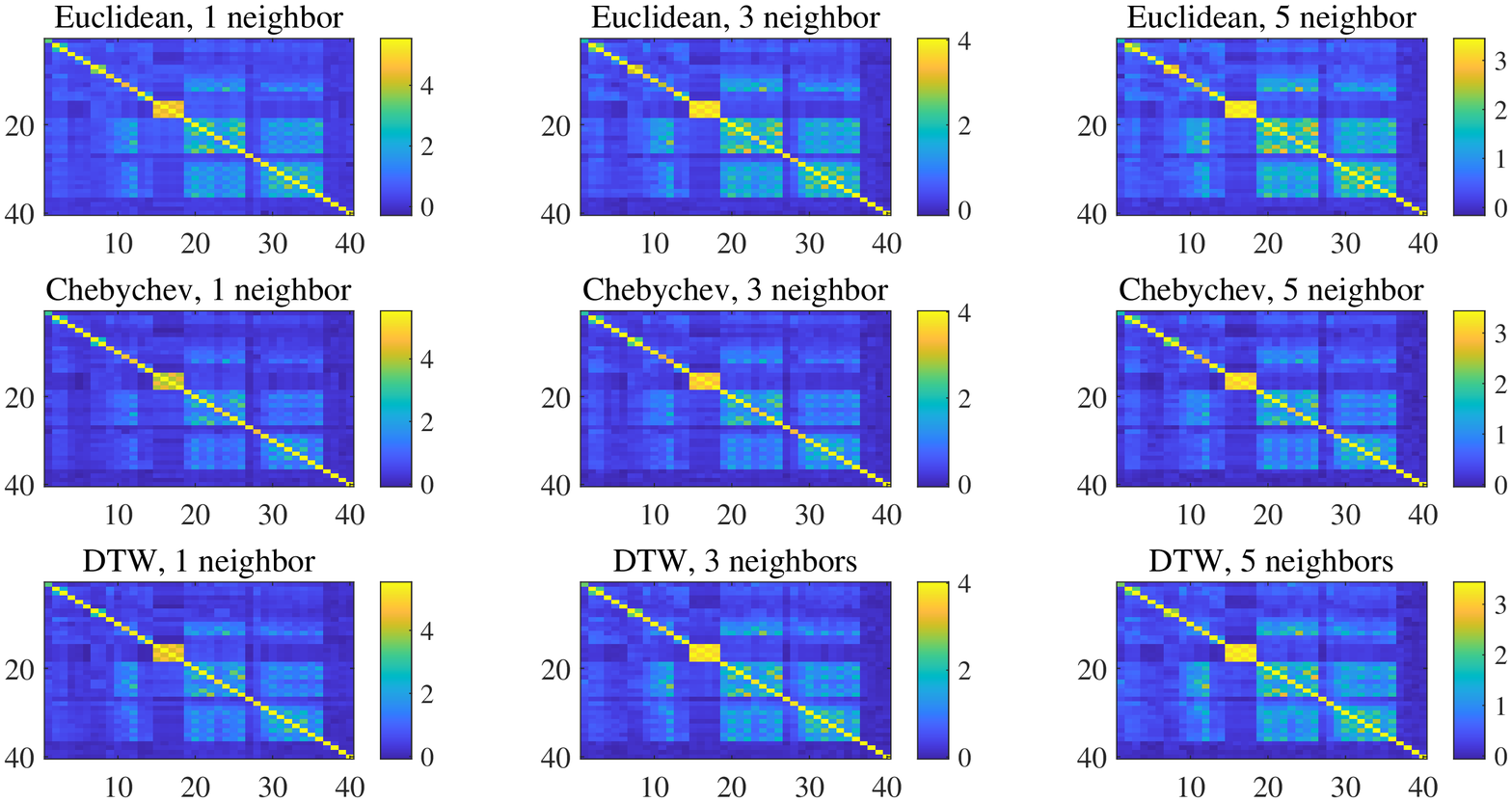}
\caption{Mutual dependency between pairs of candidates based on different distances and the $k$-nearest neighbor method. Each block in the heatmaps represents the mutual dependency of the two attributes whose index is denoted on the x- and y-axes.}\label{fig:MI_TSTS}
\end{figure}

After evaluating the informativeness of individual attributes, it is natural to consider the discriminating power of a combination of these attributes. In the context of pattern recognition and machine learning, a combination of attributes tends to achieve better results. However, it is important to check whether high mutual dependency exists among the attributes using the mutual information outlined in Section \ref{MI_TSTS}. Figure \ref{fig:MI_TSTS} shows the mutual dependency of pairs of candidates based on different distance and parameter settings. The nine heatmaps are similar, showing that only a few of the attributes such as the $MA$'s and $EMA$'s are highly dependent, while most have moderate or low levels of mutual dependency.

\begin{table}[htbp]
\centering
\caption{Selected jump indicators with different distances and $k$ parameters}\label{tab:features}
\begin{tabular}{llcccccccccccccccccccccccc}
\hline\noalign{\smallskip}
    &&\multicolumn{3}{c}{Euclidean}  &&\multicolumn{3}{c}{Chebychev}  &&\multicolumn{3}{c}{DTW}     &&\multicolumn{2}{c}{Intersection}\\
    \cline{3-5}         \cline{7-9}   \cline{11-13} \cline{15-16}
&\#Neighbors      &1 &3    &5     &&1      &3       &5         &&1      &3      &5               \\
1	&	r	      &$\surd$ 	&$\surd$ 	&$\surd$ 	&	&$\surd$ 	&$\surd$ 	&$\surd$ 	&	&$\surd$ 	&$\surd$ 	&$\surd$    &&&$\surd$ 	\\
2	&	V	      &$\surd$ 	&$\surd$ 	&$\surd$ 	&	&$\surd$ 	&$\surd$ 	&$\surd$ 	&	&$\surd$ 	&$\surd$ 	&$\surd$ 	&&&$\surd$\\
3	&	K	      &$\surd$ 	&$\surd$ 	&$\surd$ 	&	&$\surd$ 	&$\surd$ 	&$\surd$ 	&	&$\surd$ 	&$\surd$ 	&$\surd$ 	&&&$\surd$\\
4	&	S	      &$\surd$ 	&$\surd$ 	&$\surd$ 	&	&$\surd$ 	&$\surd$ 	&$\surd$ 	&	&$\surd$ 	&$\surd$ 	&$\surd$ 	&&&$\surd$\\
5	&	OI	      &$\surd$ 	&$\surd$ 	&$\surd$ 	&	&$\surd$ 	&$\surd$ 	&$\surd$ 	&	&$\surd$ 	&$\surd$ 	&$\surd$ 	&&&$\surd$\\
6	&	DI	      &$\surd$ 	&$\surd$ 	&$\surd$ 	&	&$\surd$ 	&$\surd$ 	&$\surd$ 	&	&$\surd$ 	&$\surd$ 	&$\surd$ 	&&&$\surd$\\
7	&	QS	      &	        &	        &$\surd$ 	&	&$\surd$ 	&$\surd$ 	&$\surd$ 	&	&$\surd$ 	&$\surd$ 	&$\surd$ 	\\
8	&	ES	      &$\surd$ 	&$\surd$ 	&$\surd$ 	&	&$\surd$ 	&$\surd$ 	&$\surd$ 	&	&$\surd$ 	&$\surd$ 	&$\surd$ 	&&&$\surd$\\
9	&	RV	      &$\surd$ 	&$\surd$ 	&$\surd$ 	&	&$\surd$ 	&$\surd$ 	&$\surd$ 	&	&$\surd$ 	&$\surd$ 	&$\surd$ 	&&&$\surd$\\
10	&	R	      &$\surd$ 	&$\surd$ 	&$\surd$ 	&	&$\surd$ 	&$\surd$ 	&$\surd$ 	&	&$\surd$ 	&$\surd$ 	&$\surd$ 	&&&$\surd$\\
11	&	PROC(5)   &$\surd$ 	&$\surd$ 	&$\surd$ 	&	&$\surd$ 	&$\surd$ 	&$\surd$ 	&	&$\surd$ 	&$\surd$ 	&$\surd$ 	&&&$\surd$\\
12	&	PROC(20)  &$\surd$ 	&	        &	        &	&$\surd$	&$\surd$ 	&$\surd$ 	&	&$\surd$	&$\surd$    &	\\
13	&	VROC(5)	  &$\surd$ 	&$\surd$ 	&$\surd$ 	&	&$\surd$ 	&$\surd$ 	&$\surd$ 	&	&$\surd$ 	&$\surd$ 	&$\surd$ 	&&&$\surd$\\
14	&	VROC(20)  &$\surd$ 	&$\surd$ 	&$\surd$ 	&	&$\surd$ 	&$\surd$ 	&$\surd$ 	&	&$\surd$ 	&$\surd$ 	&$\surd$ 	&&&$\surd$\\
15	&	MA(5)	  & 	    &	        &$\surd$	&	&$\surd$ 	&$\surd$ 	&	        &	&$\surd$ 	&$\surd$ 	&$\surd$ 	\\
16	&	MA(20)	  &$\surd$	&$\surd$ 	&$\surd$	&   &$\surd$ 	&$\surd$ 	&$\surd$ 	&	&$\surd$ 	&$\surd$ 	&$\surd$ 	&&&$\surd$\\
17	&	EMA(5)	  &$\surd$ 	&$\surd$ 	&  	        &	&	        &	        &$\surd$ 	&	&$\surd$ 	&$\surd$ 	&$\surd$ 	\\
18	&	EMA(20)	  &$\surd$ 	&	        &         	&	&$\surd$ 	&$\surd$ 	&$\surd$ 	&	&$\surd$ 	&$\surd$ 	&$\surd$ 	\\
19	&	BIAS(5)	  &	        &	        &	        &	&$\surd$ 	&$\surd$ 	&$\surd$ 	&	&	        &$\surd$ 	&$\surd$ 	\\
20	&	BIAS(20)  &	        &	        &	        &	& 	        &$\surd$	&$\surd$ 	&	&	        &$\surd$    &$\surd$ 	\\
21	&	EBIAS(5)  &	        &	        &$\surd$ 	&	&	        &$\surd$ 	&$\surd$ 	&	&$\surd$ 	&$\surd$ 	&$\surd$ 	\\
22	&	EBIAS(20) &	        &	        &	        &	&	        &	        &	        &	& 	        &       	&$\surd$ 	\\
23	&	OSCP(5)	  & 	    &$\surd$ 	&$\surd$ 	&	&$\surd$ 	&$\surd$ 	&$\surd$ 	&	&$\surd$	&$\surd$ 	&$\surd$ 	\\
24	&	OSCP(20)  &	        & 	        &$\surd$ 	&	&$\surd$ 	&$\surd$ 	&$\surd$ 	&	&	        &          	&$\surd$ 	\\
25	&	EOSCP(5)  &	        & 	        &	        &	&	        &	        &$\surd$ 	&	&	        &$\surd$ 	&$\surd$ 	\\
26	&	EOSCP(20) &	        &$\surd$	&	        &	&	        &	        &	        &	&$\surd$	&                &	\\
27	&	ADO	      &$\surd$ 	&$\surd$ 	&$\surd$ 	&	&$\surd$ 	&$\surd$ 	&$\surd$ 	&	&$\surd$ 	&$\surd$ 	&$\surd$ 	&&&$\surd$\\
28	&	TR	      &$\surd$ 	&$\surd$ 	&$\surd$ 	&	&$\surd$ 	&$\surd$ 	&$\surd$ 	&	&$\surd$ 	&$\surd$ 	&$\surd$ 	&&&$\surd$\\
29	&	fastpctK(5)&$\surd$ & 	        &$\surd$ 	&	&$\surd$ 	&$\surd$ 	&$\surd$ 	&	&$\surd$ 	&$\surd$ 	&$\surd$ 	\\
30	&	fastpctK(20)&$\surd$&$\surd$	&	        &	&$\surd$ 	&$\surd$ 	&$\surd$ 	&	&	        &$\surd$ 	&$\surd$ 	\\
31	&	fastpctD(5)	&$\surd$&$\surd$	&	        &	&$\surd$ 	&$\surd$ 	&$\surd$ 	&	&$\surd$ 	&$\surd$ 	&$\surd$ 	\\
32	&	fastpctD(20)&	    &	        &$\surd$ 	&	&	        & 	        &$\surd$ 	&	&$\surd$ 	&$\surd$ 	&$\surd$ 	\\
33	&	spctD(5)	&$\surd$&$\surd$ 	&$\surd$ 	&	&$\surd$ 	&$\surd$ 	&$\surd$ 	&	&$\surd$ 	&$\surd$ 	&$\surd$ 	&&&$\surd$\\
34	&	spctD(20)	&$\surd$&	        &	        &	&       	&$\surd$ 	&$\surd$ 	&	&$\surd$ 	&$\surd$ 	&$\surd$ 	\\
35	&	CCI(5)	  &$\surd$ 	& 	        &$\surd$ 	&	&$\surd$ 	&$\surd$ 	&$\surd$ 	&	&$\surd$ 	&$\surd$ 	&$\surd$ 	\\
36	&	CCI(20)	  &$\surd$ 	&$\surd$ 	&$\surd$ 	&	&$\surd$ 	&$\surd$ 	&$\surd$ 	&	&$\surd$ 	&$\surd$ 	&$\surd$ 	&&&$\surd$\\
37	&	PVT	      &$\surd$ 	&$\surd$ 	&$\surd$ 	&	&$\surd$ 	&$\surd$ 	&$\surd$ 	&	&$\surd$ 	&$\surd$ 	&$\surd$ 	&&&$\surd$\\
38	&	OBV	      &$\surd$ 	&$\surd$ 	&$\surd$ 	&	&$\surd$ 	&$\surd$ 	&$\surd$ 	&	&$\surd$ 	&$\surd$ 	&$\surd$ 	&&&$\surd$\\
39	&	NVI	      &$\surd$ 	&$\surd$ 	&$\surd$ 	&	&$\surd$ 	&$\surd$ 	&$\surd$ 	&	&$\surd$ 	&$\surd$ 	&$\surd$ 	&&&$\surd$\\
40	&	PVI	      &$\surd$ 	&$\surd$ 	&$\surd$ 	&	&$\surd$ 	&$\surd$ 	&$\surd$ 	&	&$\surd$ 	&$\surd$ 	&$\surd$ 	&&&$\surd$\\
\noalign{\smallskip} \cline{3-5}         \cline{7-9}   \cline{11-13} \cline{15-16}\noalign{\smallskip}
&Total       &29 &26 &29 &   &32	& 35	&37		&&34	&37	&38 &&&21\\
\noalign{\smallskip}\hline
\end{tabular}
\begin{tablenotes}
\item The marker $\surd$ denotes the selected attributes in different cases. The last column shows the attributes that are covered by all nine sets.
\end{tablenotes}
\end{table}

The next task is to select a combination of attributes, called jump indicators, that are highly related to the occurrence of price jumps. Because the abnormality of some attributes starts early, as concluded from Table \ref{tab:changepower}, the FSS technique, as presented in Section \ref{mRMRFS}, is performed on the set of attributes extracted within the prior 4-hour window. For comparison purposes, Table \ref{tab:features} shows the nine sets of selected jump indicators based on different types of distances and $k$ values. All the sets of the jump indicators include more than 65\% of the candidate attributes, while the Chebychev and DTW distances sometimes even select most of the jump attributes. Up to 21 attributes are included in all nine sets. Although some attributes have low informativeness such as $ADO$, $OBV$, and $PVI$, they play a role in the jump indicator set. Further, even the $MA$ and $EMA$ are highly dependent, they are simultaneously selected in most cases, indicating that complementing information can still be found in these attributes.

\subsection{Stock clustering according to the jump indicator patterns}
To cluster the stocks according to their trading abnormality before price jumps, the 189 jumping samples (i.e., the samples $(X_i, C_i)$ with $C_i=1$), as obtained in Section \ref{stockprep}, are adopted, each of which corresponds to a stock. To have a more robust clustering result with respect to the different distances, the 21 jump indicators covered by all nine sets in Table \ref{tab:features} are used to represent each sample. Thus, each $X_i$ is now represented by 21 time series of length 48. To minimize the effect of the scale difference, each of the indicator series is normalized using the min-max method.

\begin{figure}[htbp]
\begin{center}
\includegraphics[scale=0.4]{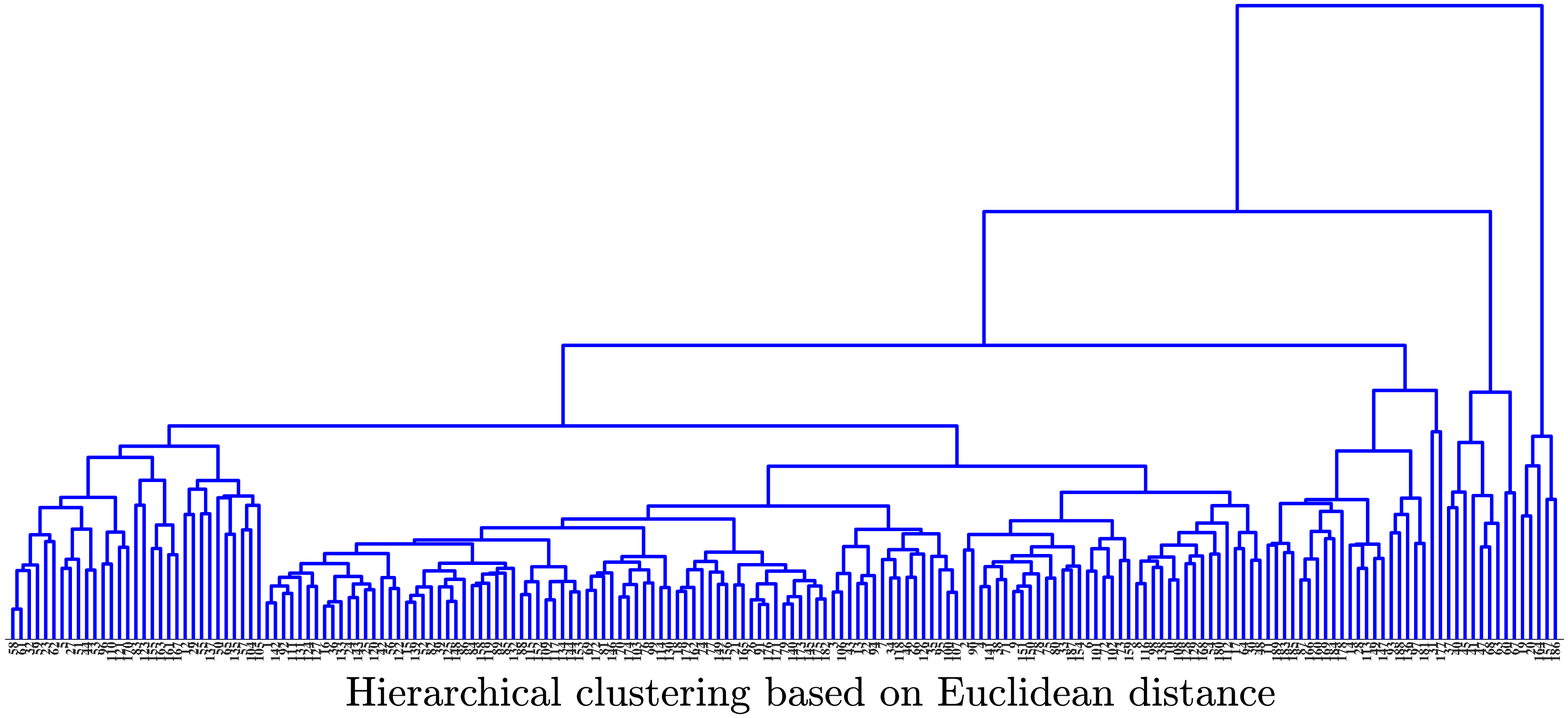}
\includegraphics[scale=0.4]{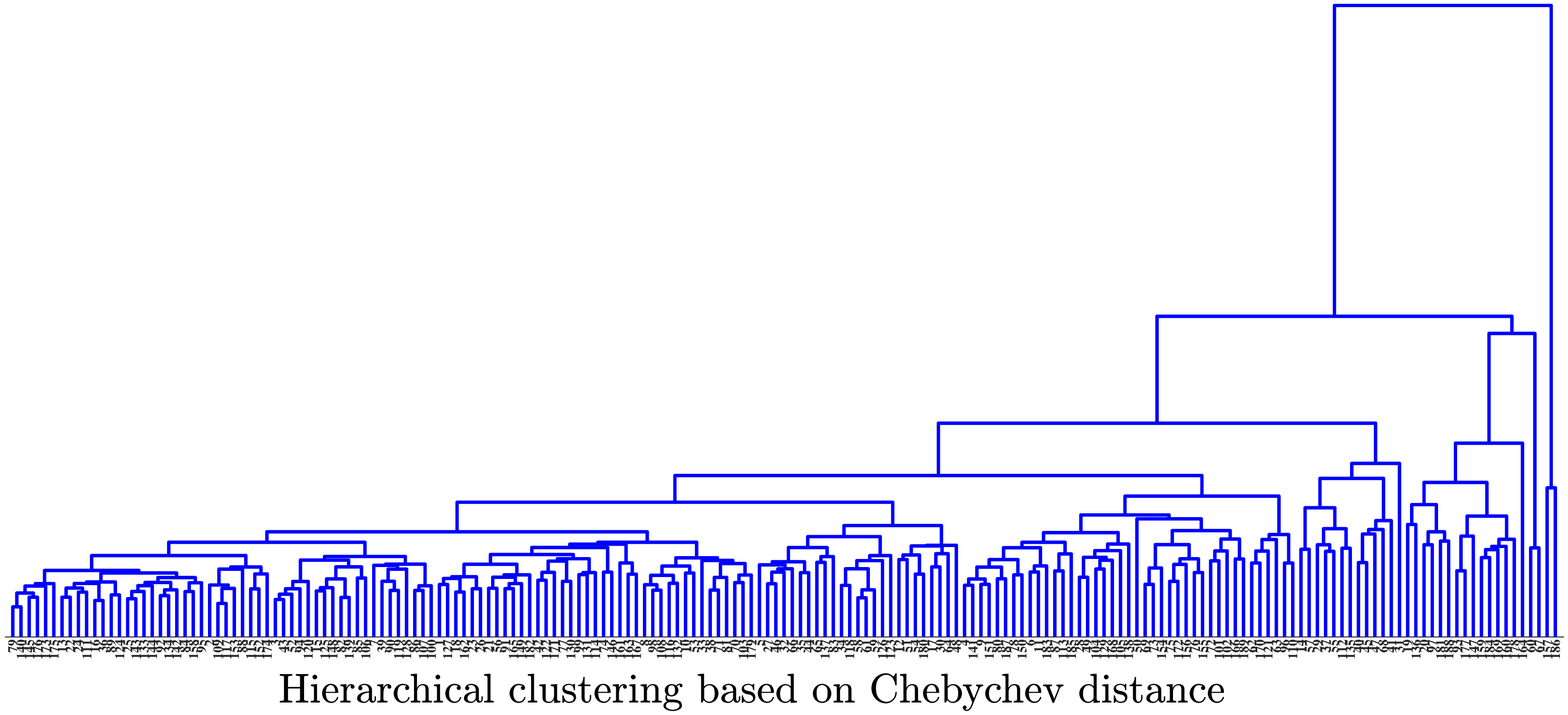}
\includegraphics[scale=0.4]{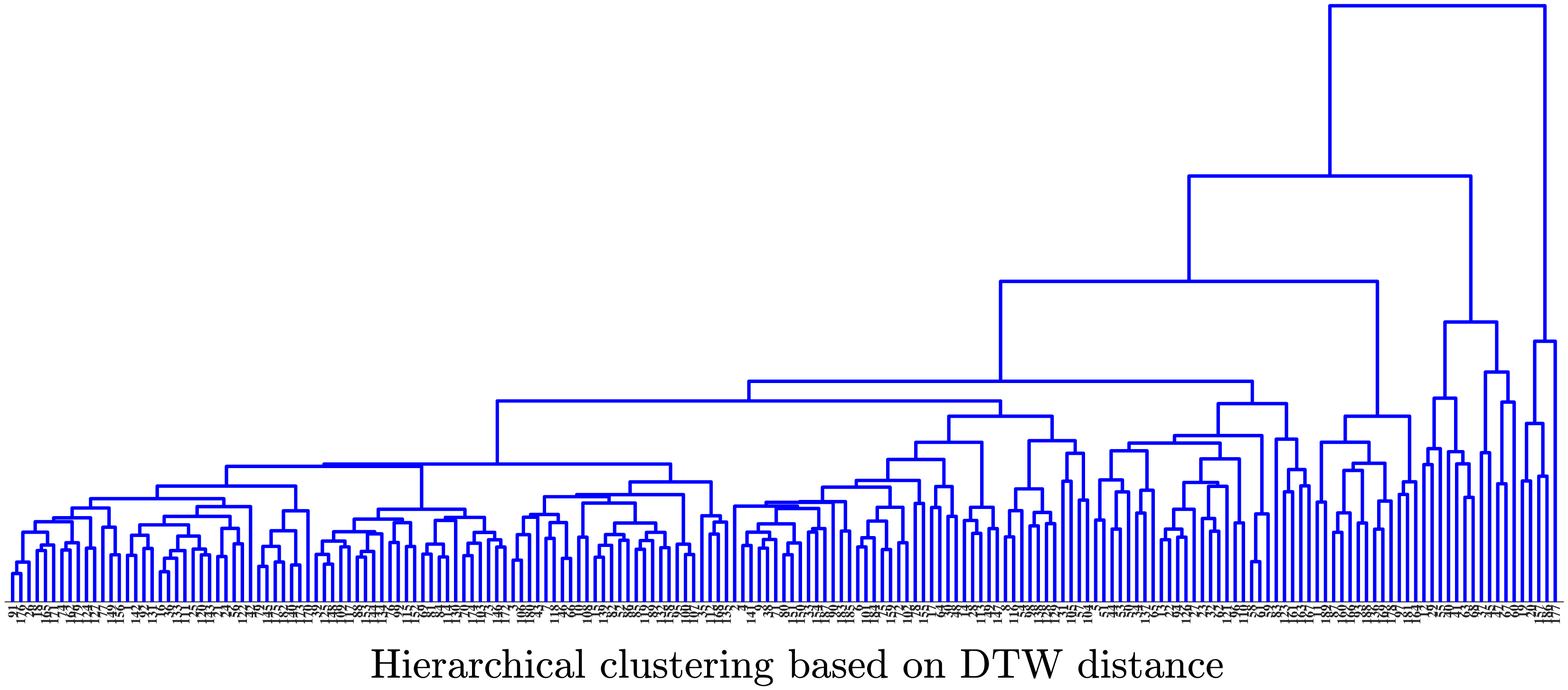}
\caption{Hierarchical clustering of the stocks}
\end{center}\label{fig:clust}
\end{figure}

Figure \ref{fig:clust} shows the hierarchical clustering results of the stocks based on the three types of distances. In the figures, the objects are linked by upside-down U-shaped lines. The height of the ``U'' represents the distance between the two clusters it links. If a link is approximately the same height as the links below it, it is said to exhibit a high level of consistency, but be inconsistent with the links below it. In a cluster analysis, inconsistent links can provide the border of a natural division in a sample set. Hence, from the following dendrogram, we can easily find the individual stocks that different from the majority.

\begin{table}[htbp]
\centering
\caption{The most distinct stocks detected by the clustering analysis}\label{tab:distinctstocks}
\begin{tabular}{ccc|cccccccccc}
\hline\noalign{\smallskip}
Euclidean  &Chebychev  &DTW     &All    &Codes\\
\noalign{\smallskip}\hline\noalign{\smallskip}
19          &157        &19     &19     &'000709'\\
20          &186        &20     &20     &'000725'\\
157         &           &157    &157    &'601288'\\
164         &           &177    &164    &'601398'\\
186         &           &186    &177    &'601800'\\
            &           &       &186    &'601988'\\
\noalign{\smallskip}\hline
\end{tabular}
\begin{tablenotes}
\item The fourth column summarizes the distinct stocks detected by all the three distances, while the fifth column provides their codes.
\end{tablenotes}
\end{table}

In these figures, the most distinct stocks are in the cluster to the right of the trees. Comparing the height of the top link to the below ones in each figure, we find that the clustering analysis based on the Chebychev distance recognizes a more different cluster to the majority than the other two types of clustering. Table \ref{tab:distinctstocks} shows the index of the most distinct stocks detected by the three clustering analyses. Compared with the Chebychev distance, the Euclidean and DTW distances detect larger and similar sets of distinct stocks. Nevertheless, the stocks detected by the Chebychev distance are included in the other two sets. It is likely because that the Chebychev distance focuses on the maximal difference but ignores the small ones between time series; in this case, the Euclidean and DTW distances can capture more features of distinct stocks when clustering, resulting in a larger set of distinct stocks.

\begin{figure}[htbp]
\begin{center}
\includegraphics[scale=0.40]{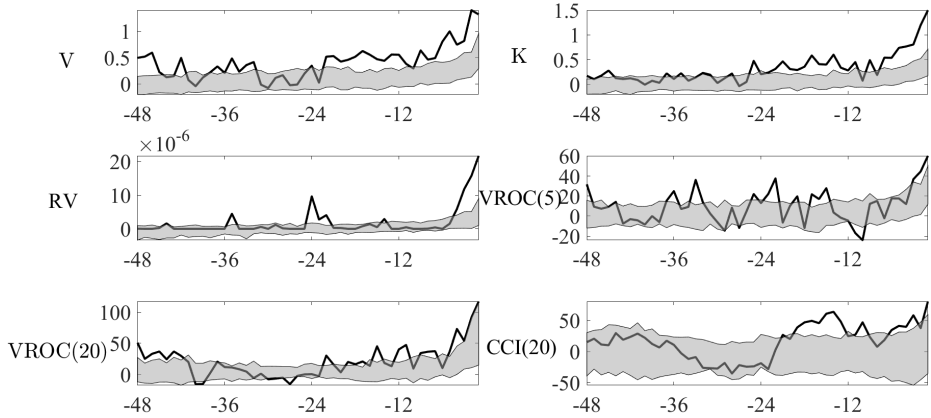}
\caption{The trading patterns of a distinct stock (stock 157) before price jumps. The solid line is the median time series of the corresponding attribute of the stock. The shaded region represents the range between the 5\% and 95\% quantiles of the median time series of the stocks absent from Table \ref{tab:distinctstocks}. The x-axis is the index of the intervals before price jumps.}
\end{center}\label{fig:distinct}
\end{figure}

Figure \ref{fig:distinct} provides, as an example, the attribute patterns of a stock (indexed by 157) before its price jumps, showing the extent to which they differ from those of most stocks. For parsimony, we only present the dynamic patterns of the six attributes that have the highest discriminating power. In the eight plots, the solid lines represent the median attribute series of stock 157 before price jumps, while the shaded regions correspond to the range between the 5\% and 95\% quantiles of the median attribute series of the stocks absent from Table \ref{tab:distinctstocks}. Compared with most stocks, the abnormal movements of the attributes $V$, $K$ and $CCI(20)$ for stock 157 start much earlier (about 2 hours, or 24 intervals) before the occurrence of price jumps; meanwhile, their abnormality is strong, as shown by the high levels of the solid lines equal to or above the 5\% quantile in their corresponding plots. On the contrary, the attributes $RV$, $VROC(5)$, and $VROC(20)$ predict the occurrence of price jumps at a slower pace: they all reach high levels, but only within 30 minutes (six intervals). Nevertheless, the strong abnormality of these attributes explains why this stock is identified as distinct by all the three clustering analyses from the perspective of micro-trading behavior before price jumps.

\section{Discussion}
In this study, we introduce a new framework based on multivariate time series analysis to investigate abnormal trading patterns before price jumps. Different from the existing literature, our methodology can help explore the temporal information in the candidate attributes used to describe trading behaviors and select a set of attributes, called jump indicators, to help recognize jump-related abnormal patterns of different stocks. In addition to commonly used liquidity measures, technical indicators are further included in the candidate attributes to describe micro-trading behaviors from a different perspective.

The empirical study is conducted using level-2 data of the constituent stocks of the CSI 300. After sample preparation, each stock is represented by a list of attribute series in two scenarios: before intraday price jumps and on steady days. Using time series-based mutual information, we find that some volume- and volatility-related attributes exhibit the most significant abnormality before price jumps, which is consistent with previous findings. The choice of distance and the $k$ parameter in the mutual information computation tend to have a low influence when comparing the informativeness of different attributes. Evaluating their mutual information values within shrinking windows, we also find that although a large number of attributes show abnormal movements shortly before the occurrence of price jumps, some still start to move abnormally much earlier. Further, the mutual information between time series shows that the attributes have low mutual dependency, which suggests various perspectives to assess micro-trading behaviors. Based on that, we then select a set of jump indicators and have detected some stocks with extremely abnormal trading behaviors before price jumps using clustering analysis. The proposed methodology as well as the empirical results obtained in our study complement existing work and provide a novel framework and perspective from which to investigate micro-trading behaviors before price jumps in financial markets.

Future research could focus on the common and idiosyncratic micro-trading patterns of individual stocks and investigate the predictability of their price jumps using the jump indicators we selected. High-frequency portfolio allocation and risk management strategies related to the occurrence of stock price jumps could also be designed.

\section*{Declarations}
\paragraph{Funding} This work was supported by the Natural Science Foundation of China (71720107001, U1811462) and the Humanities and Social Science Fund of the Ministry of Education (17YJA790101).
\paragraph{Conflicts of interest/Competing interests} The authors declared that they have no conflicts of interest to this work.
\paragraph{Availability of data and material} The data that support the findings of this study are openly available in the Wind database(http://www.wind.com.cn) provided by Shanghai Wind Information Co., Ltd.
\paragraph{Code availability}The code is available upon request.
\paragraph{Authors' contributions}All authors contributed to the study conception and design. All authors read and approved the final manuscript.
\bibliography{mybib}
\bibliographystyle{spbasic}
\end{document}